\documentstyle[preprint,aps,tighten]{revtex}
\preprint{\vbox{\baselineskip=12pt
\rightline{CGPG-97/3-12}
\rightline{hep-th/9703177}}}

\def\E{\tilde{E}}
\def\U{{\rm U}}
\def\SU{{\rm SU}}
\def\e{\varepsilon}
\def\ut#1{\rlap{\lower1ex\hbox{$\sim$}}{#1}}
\def\ket#1{|\,{#1}\,\rangle}
\def\be{\nopagebreak[3]\begin{equation}}
\def\ee{\end{equation}}
\def\ba{\nopagebreak[3]\begin{eqnarray}}
\def\ea{\end{eqnarray}}

\def\d{{\rm d}}
\def\g{\gamma}
\newcommand{\teta}{\rlap{\lower2ex\hbox{$\,\tilde{}$}}\eta{}}

\begin{document}
\draft
\title{Loop Quantization of Maxwell Theory and \\
Electric Charge Quantization}
\author {Alejandro Corichi\thanks{Electronic address: 
corichi@phys.psu.edu}
and Kirill V.\ Krasnov\thanks{Electronic address:
 krasnov@phys.psu.edu}
}
\address{Center for Gravitational Physics and Geometry \\
Department of Physics, The Pennsylvania State University \\
University Park, PA 16802, USA}
\maketitle

\begin{abstract}

We consider the loop quantization of Maxwell theory. A quantization of this 
type leads to a quantum theory in which the fundamental excitations are 
loop-like rather than particle-like. Each such loop plays the role of a 
quantized Faraday's line of electric flux. We find that the quantization 
depends on an arbitrary choice of a parameter $\e$ that carries the 
dimension of electric charge. For each value of $\e$ an electric charge 
that can be contained inside a bounded spatial region is automatically
quantized in  units of $\hbar/4\pi \e$. The requirement of consistency 
with the quantization of electric charge observed in our Universe  fixes 
a value of the, so far arbitrary, parameter $\e$ of the theory. Finally, 
we compare the ambiguity in the choice of parameter $\e$ with the 
$\beta$-ambiguity that, as pointed by Immirzi, arises in the loop 
quantization of general relativity, and comment on a possible way this
ambiguity can be fixed.

\end{abstract}

The introduction by Rovelli and Smolin of Wilson loop observables to
quantum gravity \cite{RS1} has led to the development of a new type of 
quantization for theories of connections 
\cite{AI,ALMMT}. A quantization of this kind, unlike the Fock-type
one utilized in standard quantum field theories, is independent
on any background structure as that, for example, of the Minkowski metric
on space-time.
The fundamental excitations of such quantum theories are 1-dimensional,
loop-like, rather than 3-dimensional, particle-like. 
Loops play the role of quantized flux tubes: 
for example, in quantum gravity surfaces acquire area
through their intersections with these loops. Each loop, labeled with spin
$j$, contributes an area equal to $\sqrt{j(j+1)}$ times the Planck area 
for each transverse intersection \cite{RS2,AL}. Thus, in this type
of quantization of gauge theories  Faraday's idea of {\it lines of 
force} is realized in a rather explicit fashion.

In this letter we discuss the implications of the loop quantization of  
Maxwell theory on the  quantization of electric charge.

The quantization of the free Maxwell theory that considers the Wilson loop 
functionals of the electro-magnetic potential 
was first studied by Gambini and Trias in \cite{GT}. 
Since then, canonical quantization of Maxwell theory  
using loop variables has been the subject of  attention of many authors 
(see, for example \cite{GP-book} and references therein, and \cite{aa:ac}).
Most of this research was
concentrated, however, on the use of loop variables and their 
Fock space representation (see, however, \cite{GP} and \cite{K}).
In this letter we shall be concerned with a different quantization procedure.
Namely, we consider a natural quantization
coming from a choice of a non-canonical algebra of observables in
Maxwell theory.

Let us concentrate on the kinematics of the free Maxwell theory in the 
Hamiltonian framework. Let $\Sigma$ denote the spatial hypersurface 
(which for technical reasons is supposed to be a smooth manifold).
Note that we do not assume any background structure on $\Sigma$, and our
entire discussion is therefore, diffeomorphism invariant.
Let $A_a$ be the electro-magnetic potential (it carries the dimension of 
$\sqrt{M/L}$), and let $\E^a$ be the electric field (of the dimension
$\sqrt{M/L^3}$) that plays the role of
the canonically conjugate variable.  Here `tilde' over the symbol of 
electric field indicates that it is a densitized vector field.

The phase space of the theory consists of pairs $(A_a,\E^a)$ satisfying 
appropriate fall-off conditions at the spatial infinity in the 
case of a non-compact $\Sigma$. The Poisson bracket between the canonically
conjugate variables is given by
\begin{equation}
\left \{ A_b(x), \E^a(y)\right \} = 
\,\delta_b^a\,{\delta}^3(x,y),
\end{equation}
where ${\delta}^3(x,y)$ stands for the $\delta$-function
defined so that $\int_\Sigma {\delta}^3(x,y) f(y) d^3y = f(x).$

The electro-magnetic potential has  a dimension different from the standard
dimension $1/L$ of a connection field. In order to convert the 
electro-magnetic potential into a $\U(1)$ connection field
one has to introduce a dimensionfull parameter into the theory.
A possible way to do this is
to introduce a parameter $\e$ that carries the dimension of electric charge
$[\e]=\sqrt{ML}$. Then $A_a/\e$ has the dimension of a connection.
 Thus, the set of fields $\frac{i}{\e}\,A_a$ becomes the 
set of $\U(1)$ connection fields on some $\U(1)$ bundle over $\Sigma$.
One can use $\frac{i}{\e}\,A_a$ to construct a holonomy $h_\gamma$ along a 
path $\gamma\in\Sigma$
\begin{equation}
h_\gamma:=\exp{\left({i\over \e}\int_\gamma A\right)}.
\end{equation}
Since the connection $\frac{i}{\e} A_a$ is abelian we use 
the ordinary exponential instead of a path ordered one.
Note that so far $\e$ is an arbitrary parameter with the dimension of
electric charge.

Let us now turn to the quantum theory. The algebra of observables we
want to regard as fundamental in the quantization procedure is the
one generated by holonomies $h_\g$ of the $\U(1)$ connection as
configuration observables, and fluxes of electric field $E[S]$
as momenta. Recall that since $\E^a$ is a vector density of weight one,
there is a naturally defined two-form $E_{ab}$ associated to it:
$E_{ab}:=\teta_{abc}\E^c\label{efield}$.
In the same way that connection 1-forms are objects that one 
can naturally integrate  along
loops  (in order to define a holonomy $h_\g$), one can naturally
integrate 2-forms over  surfaces. Therefore, the functions 
$E[S]:=\int_S E_{ab}\d\sigma^{ab}$ are the corresponding momenta 
observables. They satisfy the following Poisson bracket relations,
\be
\Big\{ h_{\g}, E[S]\Big\} =\frac{i}{\e}\,h_{\g}\,I(\g,S)\,,\label{PB2}
\ee
where $I(\g,S)$ denotes the oriented intersection number between
the loop $\g$ and the surface $S$. 
This corresponds to the `loop-surface algebra' of ref. \cite{AI2}. 
It is important to note that there is a 1-parameter family of such
algebras, labeled precisely by the parameter $\e$.

The so called loop quantization of a gauge theory is constructed
by taking the traced holonomies 
of the connection as main configuration observables
that determine the quantum representation \cite{AI}. The kinematics of
the resultant quantum theory is described in details in \cite{ALMMT}.
For our purposes it is sufficient to recall that there is a basis of
quantum states given by spin networks, i.e., graphs embedded in $\Sigma$
with edges labeled by irreducible representations of the gauge group and the
vertices labeled by intertwining operators. In our case the gauge group is 
$\U(1)$ and irreducible representations are in one to one correspondence with
integers $q$ (called `charges'). Given a vertex of a spin network 
and a set of incoming
and outgoing edges, the intertwining operator exists if the sum of charges 
labeling the incoming edges is equal to the sum of charges labeling
the outgoing edges. In this case the intertwining operator is
unique up to a multiplicative constant factor. 

We shall now see that the edges of spin networks play the role of quantized
flux tubes of electric field. The smeared electric field observable
$E[S]$ becomes an operator in the  quantum theory. First, let us
consider the canonical commutation relations between 
the fundamental variables,
\begin{equation}
\left[ \hat{A}_b(x),\hat{E^a}(y) \right] = i\hbar
\,\delta_b^a\,{\delta}^3(x,y).
\end{equation}
The standard way to satisfy the commutation relations is to represent
$\hat{E}^a(x)$, heuristically,  
as a functional derivative with respect to $A_a(x)$ 
\begin{equation}
\hat{E^a}(x):={\hbar\over i}\,{\delta\over\delta A_a(x)}
\end{equation}

The operator $\hat{E}[S]$ then can be promoted into a well-defined
operator in the Hilbert space using the regularization technique
developed in \cite{AL}. The resulting operator 
is diagonal in the basis formed by spin network states
\begin{equation}
\hat{E}[S]\cdot\ket{\Psi} = {\hbar\over \e}\,
\sum_v\,{1\over 2} (q_v^{(u)}-q_v^{(d)})\,\ket{\Psi},
\label{qcharge}
\end{equation}
where the sum on the right hand side is taken over all vertices $v$ of the
spin network $\Psi$ lying on the surface $S$, and $q_v^{(u)}, q_v^{(d)}$ are
the sum of all charges labeling edges lying up and down the surface $S$
respectively. Here we assume that some orientation of the surface $S$
is chosen, and that all edges intersecting $S$ from below are 
oriented towards the `interior' of $S$, all edges intersecting $S$ from above
are oriented outwards. In other words, the orientation of the edges is such
that all the edges are outgoing.

In the case when all vertices of $\Psi$ lying on $S$ are bi-valent (i.e.,
those coming from simple intersections of edges of $\Psi$ with $S$), the
formula (\ref{qcharge}) simplifies
\begin{equation}
\hat{E}[S]\ket{\Psi} = {\hbar\over \e}\,\sum_v\,q_v\,\ket{\Psi},
\label{qcharge1}
\end{equation}
where $q_v$ are the charges labeling edges of $\Psi$ intersecting $S$.
It is straightforward to check that the operators thus defined
satisfy the quantum algebra coming from (\ref{PB2}). 

Thus, in our quantum theory the edges of spin networks indeed play the role of
the quantized flux tubes of electric field: flux of electric field 
through $S$ acquires value via intersections with these edges, each edge
labeled with charge $q$ contributing a flux equal to $q$ times $\hbar/\e$
for each transverse intersection. 

Let us now consider the operator of electric charge.
According to the Gauss's law, the total electric charge contained inside
a {\it closed} surface $S$ in $\Sigma$ is determined by the flux of electric 
field through $S$
\begin{equation}
Q_{S} = {1\over 4\pi}\,E[S],
\end{equation}
where $S$ is a closed surface.
The corresponding quantum operator $\hat{Q}_{S}$
is given by ($\frac{1}{4\pi}$ times) Eq. (\ref{qcharge}).

Let us analyze the spectrum of $\hat{Q}_S$. In the free Maxwell theory, which
we were considering so far, a charge contained inside a closed surface $S$
in $\Sigma$ is zero when the topology of $\Sigma$ is trivial. It is 
interesting to note, however, that even in the free Maxwell theory one
can have a non-zero charge inside a closed surface in the case of a non-trivial
topology of the spatial manifold.  This possibility, first pointed out 
by Misner and Wheeler \cite{wheeler}, is fully realized in our theory.

To have a possibility of a non-zero charge inside a closed surface in the 
case of the trivial topology of $\Sigma$ one has to include a charged matter
into the theory. The most natural possibility would be to couple the
theory to fermionic matter. The corresponding quantum theory has been
constructed \cite{K,BK}.  However, we will not need the 
details of that construction here. What is important for us in this
letter is that, in the case a fermionic charged matter is present in
the theory, the eigenvalues of the flux operator $\hat{E}[S]$, $S$ being
closed, are all eigenvalues one finds in Eqs. (\ref{qcharge}).
Thus, when  charged matter is present in the theory, the electric charge
inside a closed surface $S$ is not necessarily zero.

Therefore, as it can be seen from (\ref{qcharge1}), 
the electric charge that can be
contained inside a closed surface $S$ is quantized in our theory
in the units of charge $\bar{e}$
\begin{equation}
\bar{e} = {1\over 4\pi}\,{\hbar\over \e}.
\end{equation}
This holds both for the case of trivial and non-trivial topologies of 
$\Sigma$. In the case of the trivial topology the charge is quantized
in the units of $\bar{e}$ when  charged matter is present. In the
case of a non-trivial topology of $\Sigma$ charge contained inside
a closed surface can be non-zero even in the free Maxwell theory.
In this case charge is again quantized in the units of $\bar{e}$.
Thus, contrary to what one would naively expect, the electric charge
is quantized not in the units of charge $\e$, but in the units of some 
different charge $\bar{e}$.

Therefore, the quantization of the theory of the type adopted here 
automatically guarantees that the electric charge is quantized. Note, however,
that in Nature the electric charge is known to be quantized in the 
units of the charge $e$ of an electron (or, as in the standard model
of elementary particles, in units of $e/3$). So the requirement that
the theory is consistent with observations must fix $\bar{e}=e$
(or, possibly, $\bar{e}=e/3$). For the discussion
that follows let us write $\bar{e}=e/n$. Then $n=1$ corresponds to 
a Universe in which electric charge is quantized in the units of the charge of
an electron, $n=3$ corresponds to a Universe in which the smallest 
quantum of electric charge is that of a quark. Then, recalling that
$(e)^2/\hbar = \alpha$ is the fine structure constant, which in our 
Universe is known to be approximately equal to $1/137$, we find that
the parameter of the theory $\e$ should be related to the charge of an 
electron in the following way
\begin{equation}
\e = {n\over 4\pi\alpha} e.
\end{equation}
Interestingly, already for $n=1$ the dimensionless quantity 
$\e^2/\hbar$ has the value $\e^2/\hbar\approx 0.87$; 
that is, the parameter of the 
theory $\e$ is required to be much larger than the electron charge.

To summarize, let us say that the loop quantization of Maxwell theory
depends on an arbitrary choice of the parameter $\e$ having the dimension of
electric charge. The quantum theory, however, predicts that the charge
is quantized in the units of $\hbar/4\pi \e$, and the requirement of 
consistency with the quantum of charge that is observed in Nature determines 
$\e$ unambiguously. This is the first result of this letter.

It is instructive to compare the results we have obtained with the
analogous results known from the loop representation of quantum gravity.
As we have mentioned above, the loop quantization of gravity predicts that 
the area of surfaces is quantized. The edges of spin networks here play
the role of quantized flux tubes of area: each intersection of an edge
labeled with spin $j$ with a surface contributes $\sqrt{j(j+1)}$ times the
Planckian length squared to the area of that surface. However, as it
was first pointed out by Immirzi in \cite{I}, there is an ambiguity in the
definition of a $\SU(2)$ connection field $A_a^i$ from the geometrical
variables (see also \cite{RT}).
Possible choices of the connection field $A_a^i$ are labeled
by a real parameter $\beta$. Starting from different $A_a^i$ one gets 
non-equivalent quantum theories. In particular, the eigenvalues of operators
measuring areas of surfaces happen to depend on a choice of $\beta$  
\begin{equation}
A_S = \beta\,l_p^2\,\sum_v \sqrt{j_v(j_v+1)}.
\label{qarea}
\end{equation}
Here the sum is taken over all edges intersecting $S$ transversally,
and all intersections are assumed to be simple bi-valent.

The above $\beta$-ambiguity in quantum gravity is quite reminiscent to the 
ambiguity in a choice of parameter $\e$ in the loop quantization of 
Maxwell theory. As we have seen, the eigenvalues of operators measuring 
charge in a region of space depend on a choice of $\e$, just as
in loop quantum gravity the eigenvalues of area operators depend on a 
choice of $\beta$. In the case of Maxwell theory the problem  
can be resolved by a requirement of consistency with the observed quantization
of electric charge. In the case of quantum gravity, however, one is
unable to fix the $\beta$-ambiguity in such a way, since there is no 
way the predicted spectrum of area operators can be compared with 
experiment because of the tiny value of $l_p^2$ as compared with our
ordinary scale.

Let us conclude this note by pointing out a possible way 
the $\beta$-ambiguity in quantum
gravity can be fixed. Since a direct experimental check of the spectrum
(\ref{qarea}) does not seem to be possible nowadays, 
one would have to find a macro-scale
consequence of the theory that depends on the detailed form of the
area spectrum. Thus, comparing such a prediction of the 
quantum theory with experiment, or with results predicted by other
well-established theories, one could fix the value of $\beta$
from the requirement of consistency. As such a result of the quantum
theory one can take, for example, the recent statistical mechanical
calculations of black hole entropy based on the formalism of loop 
quantum gravity \cite{KR}. 
These calculations yield for the statistical mechanical
entropy of Schwarzschild black hole $S=cA/\beta l_p^2$, where $c$ is
the dimensionless constant determined by calculations. A comparison
of this result
with the Bekenstein-Hawking entropy $S=A/4 l_p^2$  can be used to determine
a value of parameter $\beta$. 
This provides  one with a possible way
the parameter $\beta$ can be fixed in quantum gravity. This is the second
observation of this letter.

The authors would like to thank all the participants of the Penn State
Mini-workshop on Quantum Gravity, February 1997, for stimulating
discussions. We are particularly grateful to
A. Ashtekar, J. Baez, D. Marolf, C. Rovelli and T. Thiemann for
a discussion that clarified the analogy between the $\e$-ambiguity
in Maxwell theory and the  $\beta$-ambiguity in quantum gravity.
AC was supported by Universidad Nacional
Aut\'onoma de M\'exico (DGAPA, UNAM). This work was in part supported by
NSF-grant PHYS 95-14240 and the Eberly reserch fund of Penn State.

\end{document}